\documentclass{ws-ijmpa}

\begin{document}

\markboth{J. Vijande}{}

%
\catchline{}{}{}{}{}
%

\title{Understanding open-charm mesons}

\author{\footnotesize J. Vijande$^{1,2}$\footnote{E-mail address: javier.vijande@uv.es},
F. Fern\'andez$^1$, A. Valcarce$^1$.
}

\address{
$^{1}$Grupo de F{\'\i}sica Nuclear and IUFFyM,
Universidad de Salamanca, E-37008 Salamanca, Spain\\
$^{2}$Dpto. de F\' \i sica Te\'orica and IFIC,
Universidad de Valencia - CSIC,
E-46100 Burjassot, Valencia, Spain\\
}

\maketitle


\begin{abstract}
We present a theoretical framework that accounts for
the new $D_J$ and $D_{sJ}$ mesons measured in
the open-charm sector. These resonances are
properly described if considered as a mixture
of conventional $P-$wave quark-antiquark
states and four-quark components.
The narrowest states are basically $P-$wave quark-antiquark
mesons, while the dominantly four-quark states are shifted above
the corresponding two-meson threshold.
We study the electromagnetic decay widths
as basic tools to scrutiny their nature.
\end{abstract}

\vspace{0.5cm}

During the last few years, heavy meson spectroscopy is living a 
continuous excitation due to the discovery of several new charmed 
mesons. Three years ago BABAR Collaboration reported the observation 
of a charm-strange state, the $D_{sJ}^*(2317)$\cite{Bab03}, that was later on confirmed by
CLEO\cite{Cle03} and Belle Collaborations\cite{Bel04}.
Besides, BABAR had also pointed out to the existence 
of another charm-strange meson, the $D_{sJ}(2460)$\cite{Bab03}. This
resonance was measured by CLEO\cite{Cle03} and 
confirmed by Belle\cite{Bel04}. 
Belle results are consistent with the 
assignments of $J^P=0^+$ for the 
$D^*_{sJ}(2317)$ and $J^P=1^+$ for the $D_{sJ}(2460)$.
However, although these states are well established,
they present unexpected properties quite different from
those predicted by quark potential models. If they would correspond 
to standard $P-$wave mesons made of a charm quark, $c$, and a
strange antiquark, $\overline{s}$, their masses would be larger,
around 2.48 GeV for the $D_{sJ}^*(2317)$ and 2.55 GeV 
for the $D_{sJ}(2460)$. They would be therefore above the
$DK$ and $D^*K$ thresholds, respectively, being broad resonances.
However the states observed by BABAR and CLEO are very
narrow, $\Gamma < 4.6$ MeV for the $D_{sJ}^*(2317)$ 
and $\Gamma < 5.5$ MeV for the $D_{sJ}(2460)$. 

The intriguing situation of the charm-strange mesons 
has been translated to the nonstrange sector with the Belle 
observation\cite{Belb4} of a nonstrange broad scalar resonance, $D^*_0$,
with a mass of $2308\pm 17\pm 15\pm 28$ MeV/c$^2$
and a width $\Gamma=276 \pm 21 \pm 18 \pm 60$ MeV.
A state with similar properties has been suggested by 
FOCUS Collaboration at Fermilab\cite{Foc04} during the 
measurement of masses and widths of excited charm mesons
$D^*_2$. This state generates 
for the open-charm nonstrange mesons a very similar
problem to the one arising in the strange sector with the
$D_{sJ}^*(2317)$. If the $D^*_0(2308)$  would correspond 
to a standard $P-$wave meson made of a charm quark, $c$, and a
light antiquark, $\overline{n}$, its mass would have to be larger,
around 2.46 GeV. In this case, the quark potential models prediction and
the measured resonance are both above the $D\pi$ 
threshold, the large width observed being expected although not
its low mass.

The difficulties to identify the $D_J$ and $D_{sJ}$ states with 
conventional $c\overline{n}$ mesons are rather similar to those 
appearing in the light-scalar meson sector\cite{Bal03} and may be indicating
that other configurations are playing a role.
$q\overline{q}$ states are more easily identified with physical hadrons 
when virtual quark loops are not important. 
This is the case of the pseudoscalar 
and vector mesons, mainly due to the $P-$wave nature of this hadronic dressing. 
On the contrary, in the scalar sector is the $q\overline{q}$ pair the one in
a $P-$wave state, whereas quark loops may be in a $S-$wave. 
In this case the intermediate hadronic states that are created 
may play a crucial role in the composition of the resonance,
in other words unquenching is important. 
This has been shown to be relevant for the proper description 
of the low-lying scalar mesons\cite{Vij05}. 

In this work we have explored the same ideas for the
understanding of the properties of the 
$D_J$ and $D_{sJ}$ meson states. 
In non-relativistic quark models the wave 
function of a zero baryon number (B=0) hadron may be written as
$\left|\rm{B=0}\right>=\Omega_1\left|q\bar q\right>+\Omega_2\left|qq\bar q \bar q\right>+....$
where $q$ stands for quark degrees of 
freedom and the coefficients $\Omega_i$ take into 
account the mixing of four- and two-quark states.
The hamiltonian considering the mixing between
both configurations could be described using the $^{3}P_{0}$ model, however, since this 
model depends on the vertex parameter, we prefer in a
first approximation to parametrize this coefficient 
by looking to the quark pair that is annihilated and 
not to the spectator quarks that will form the final 
$q\overline{q}$ state. Therefore we have taken 
$V_{q\overline{q} \leftrightarrow qq\bar{q}\bar{q}}=\gamma $.
Further details about the formalism and the constituent 
quark model used are given in Refs.\cite{Vij05,Vijb5}. 

\begin{table}
\tbl{$c\overline s$ and $c\overline n$ masses (QM), in MeV.
Experimental data (Exp.) are taken from
Ref.\protect\cite{Eid04}, except for the state denoted by a dagger
that has been taken from Ref.\protect\cite{Belb4}.}
{\begin{tabular}{|cc|ccc||ccc|}
\hline
&$nL$ $J^P$&State		  &QM $(c\overline s)$  &Exp.				&State			&QM $(c\overline n)$ 	&Exp.\\
\hline
&$1S$ $0^-$&$D_{s}$ 	  & 1981  &1968.5$\pm$0.6	&$D$			&1883 	&1867.7$\pm$0.5\\
&$1S$ $1^-$&$D^*_{s}$	  & 2112  &2112.4$\pm$0.7	&$D^*$			&2010 	&2008.9$\pm$0.5\\
&$1P$ $0^+$&$D^*_{sJ}(2317)$& 2489  &2317.4$\pm$0.9	&$D^*_0(2308)$	&2465 	&2308$\pm$17$\pm$15$\pm$28$^\dagger$\\
&$1P$ $1^+$&$D_{sJ}(2460)$& 2578  &2459.3$\pm$1.3	&$D_1(2420)$	&2450 	&2422.2$\pm$1.8\\
&$1P$ $1^+$&$D_{s1}(2536)$& 2543  &2535.3$\pm$0.6	&$D_1^0(2430)$	&2546 	&$2427\pm26\pm25$\\
&$1P$ $2^+$&$D_{s2}(2573)$& 2582  &2572.4$\pm$1.5	&$D_2^*(2460)$	&2496 	&2459$\pm$4\\
\hline
\end{tabular} \label{t1}}
\end{table}

A thoroughly study of the full meson spectra has been presented
in Ref.\cite{Vijb5}. The results for the open-charm mesons are 
resumed in Table \ref{t1}. It can be
seen how the open-charm states are easily identified with
standard $c \overline{n}$ mesons except for the cases of
the $D_{sJ}^*(2317)$, the $D_{sJ}(2460)$, and the $D^*_0(2308)$.
This is a common behavior of almost all quark potential model 
calculations\cite{God91}. In a similar manner, quenched lattice NRQCD 
predicts for the $D_{sJ}^*(2317)$ a mass of 2.44 GeV\cite{Hei00}, while using
relativistic charm quarks the mass obtained is 2.47 GeV\cite{Boy97}.
Unquenched lattice QCD calculations of $c \overline s$ 
states do not find a window for the $D^*_{sJ}(2317)$\cite{Bal03},
supporting the difficulty of a $P-$wave $c\overline s$ interpretation. 

Using for the $qq$ interaction the parametrization of Ref.\cite{Vij05},
the results obtained for the $cn\bar s\bar n$ 
configuration are 2731 and 2699 MeV for the $J^{P}=0^{+}$ with $I=0$ and $I=1$, and 2841 and 2793
MeV for the $J^{P}=1^{+}$ with $I=0$ and $I=1$. For the $cn\bar n\bar n$ 
configuration with $I=1/2$ the energy is 2505 MeV.
The $I=1$ and $I=0$ states are far above the corresponding 
strong decaying thresholds and therefore should be broad, what rules
out a pure four-quark interpretation of the new open-charm mesons. 

As outlined above, for $P-$wave mesons the hadronic dressing is in a $S-$wave,
thus physical states may correspond to a mixing of 
two- and four-body configurations.
In the isoscalar sector, the 
$cn\bar s \bar n$ and $c\bar s$ states get mixed, as it happens
with $cn \bar n\bar n$ and $c\bar n$ for the $I=1/2$ case. 
The parameter $\gamma$ has been fixed to reproduce the 
mass of the $D_{sJ}^*(2317)$ meson, $\gamma=240$ MeV. 
The results obtained are shown in Table \ref{t3}. 
Let us first analyze the nonstrange sector. The 
$^{3}P_{0}$ $c\bar n$ pair and the $cn\bar n\bar n$ 
have a mass of 
2465 MeV and 2505 MeV, respectively. Once the mixing is considered
one obtains a state at 2241 MeV with 46\% of four-quark component 
and 53\% of $c\bar n$ pair. The lowest state, representing
the $D^*_0(2308)$, is above the isospin preserving threshold $D\pi$,
being broad as observed experimentally. 
The mixed configuration compares much better with 
the experimental data than the pure $c\bar n$ state. 
The orthogonal state appears higher in energy, at 2713 MeV, with
and important four-quark component. 

\begin{table}
\tbl{Probabilities (P), in \%, of the wave function components 
and masses (QM), in MeV, of the open-charm mesons 
once the mixing between $q\bar q$ and $qq\bar q\bar q$ configurations 
is considered. Experimental data are taken from Ref.\protect\cite{Eid04} 
except for the state denoted by a dagger that has been taken from Ref.\protect\cite{Belb4}.} 
{ \begin{tabular}{|c|c||c|cc||c|c|}
\hline
\multicolumn{5}{|c||}{$I=0$} & \multicolumn{2}{|c|}{$I=1/2$} \\
\hline
\multicolumn{2}{|c||}{$J^P=0^+$}    & \multicolumn{3}{|c||}{$J^P=1^+$} & \multicolumn{2}{|c|}{$J^P=0^+$} \\
\hline
QM                  &2339   		&QM						&2421  			&2555  			&QM                 &2241   \\
Exp.                &2317.4$\pm$0.9	&Exp.					&2459.3$\pm$1.3	&2535.3$\pm$0.6	&Exp.   			&2308$\pm$17$\pm$15$\pm$28$^\dagger$\\
\hline
P($cn\bar s\bar n$) &28   			&P($cn\bar s\bar n$)	&25  			&$\sim 1$ 		&P($cn\bar n\bar n$)&46         \\
P($c\bar s_{1^3P}$) &71   			&P($c\bar s_{1^1P}$)	&74  			&$\sim 1$ 		&P($c\bar n_{1P}$)  &53         \\
P($c\bar s_{2^3P}$) &$\sim 1$  		&P($c\bar s_{1^3P}$)	&$\sim 1$ 		&98				&P($c\bar n_{2P}$)  &$\sim 1$  \\
\hline
\end{tabular}
\label{t3}}
\end{table}

Concerning the strange sector, the $D_{sJ}^*(2317)$ and the $D_{sJ}(2460)$ 
are dominantly $c\bar s$ $J=0^+$ and $J=1^+$ states, respectively,
with almost  30\% of four-quark component. Such component is responsible 
for the shift of the mass of the unmixed states to 
the experimental values below the $DK$ and $D^*K$ thresholds.
Being both states below their isospin-preserving 
two-meson threshold, the only allowed strong decays to
$D_s^* \pi$ would violate isospin and are expected to
have small widths ${\it O}(10)$ keV\cite{Bar03,God03}.
As a consequence, they should be narrower than the $D_{s2}(2573)$ and 
$D_{s1}(2536)$, opposite to what it is expected from heavy quark symmetry.  
The second isoscalar $J^P=1^+$ state, with an energy of 2555 MeV and 
98\% of $c\bar{s}$ component, corresponds to the $D_{s1}(2536)$. 
Regarding the $D_{sJ}^*(2317)$, it has been argued that a 
possible $DK$ molecule would be preferred with
respect to an $I=0$ $cn\bar s\bar n$ tetraquark, 
what would anticipate an $I=1$ $cn\bar s\bar n$ partner 
nearby in mass\cite{Barb3}. Our results confirm the last argument, 
the vicinity of the isoscalar and isovector tetraquarks,
however, the restricted coupling to the $c\bar s$ system allowed only for the
$I=0$ four-quark states opens the possibility of a mixed nature for the
$D_{sJ}^*(2317)$ while the $I=1$ $J=0^{+}$ and $J=1^{+}$ four-quark states appear above
2700 MeV and cannot be shifted to lower energies.

Apart from the masses, the structure of the $D^*_{sJ}(2317)$ and 
the $D_{sJ}(2460)$ mesons could be scrutinied also through the study
of their electromagnetic decay widths. We compare in Table \ref{t4} 
our results with different theoretical approaches and the experimental 
limits reported by Belle and CLEO. 
The main difference is noticed in the suppression predicted 
for the $D_{sJ}(2460)\to D^{*+}_s\gamma$ decay as 
compared to the $D_{sJ}(2460)\to D^{+}_s\gamma$. 
A ratio ${D_{sJ}(2460)\to D^{+}_s\gamma}/{D_{sJ}(2460)
\to D^{*+}_s\gamma}\approx1-2$ has been obtained 
assuming a $q\bar q$ structure for both states\cite{Bar03,God03}
(what seems incompatible with their properties).
We find a larger value,
${D_{sJ}(2460)\to D^{+}_s\gamma}/{D_{sJ}(2460)
\to D^{*+}_s\gamma}\approx100$, due to the small
$1^3P_1$ $c\overline s$ probability of the 
$D_{sJ}(2460)$. A similar enhancement 
has been obtained in Ref.\cite{Col05} in the
framework of light-cone QCD sum rules.

This work has been partially funded by Ministerio de Ciencia y Tecnolog\'{\i}a
under Contract No. FPA2004-05616, by Junta de Castilla y Le\'{o}n
under Contract No. SA-104/04, and by Generalitat Valenciana under Contract No.
GV05/276.

\begin{table}
\tbl{Electromagnetic decay widths, in keV, for the $D^*_{sJ}(2317)$ 
and $D_{sJ}(2460)$ (QM), compared to the results of two different 
quark models based only on $q\overline q$ states. 
To compare with the experimental data by CLEO and 
Belle we have assumed for $\Gamma(D_s^{*+}\pi^0)\approx
\Gamma(D_s^+\pi^0)\approx10$ keV 
as estimated in Ref. \protect\cite{God03}.}{
\begin{tabular}{|c|ccc|cc|}
\hline
 & \multicolumn{3}{|c|}{Quark models} & \multicolumn{2}{|c|}{Experiments} \\
Transition & QM &Ref.\cite{Bar03} &Ref.\cite{God03} &CLEO\cite{Cle03}& Belle\cite{Bel04} \\
\hline
${D^*_{sJ}(2317)\to D^{*+}_s\gamma}$ & 1.6  & 1.74  & 1.9 & $<0.59$ 	&$<1.8$ \\
${D_{sJ}(2460)\to D^{*+}_s\gamma}$   & 0.06 & 4.66  & 5.5 & $<1.6$   	&$<3.1$\\
${D_{sJ}(2460)\to D^{+}_s\gamma}$    & 6.7  & 5.08  & 6.2 & $<4.9$ 		&5.5$\pm$1.3$\pm0.8$\\
\hline
\end{tabular} \label{t4}}
\end{table}

\end{document}